\documentclass[twocolumn,showpacs,preprintnumbers,amsmath,amssymb]{revtex4}
\usepackage{graphicx}
\usepackage{dcolumn}
\usepackage{bm}
\begin{document}
\title{The Three Extreme Value Distributions: An Introductory Review}         
\author{Alex Hansen}
\email{Alex.Hansen@ntnu.no}   
\address{PoreLab, Department of Physics, Norwegian University of Science and Technology, N--7491 Trondheim, Norway}     
\date{\today}          
\begin{abstract}
The statistical distribution of the largest value drawn from a sample of a given size has only three possible shapes: it is either a Weibull, a Fr{\'e}chet or a Gumbel extreme value distributions.  
I describe in this short review how to relate the statistical distribution followed by the numbers in the sample to the associate extreme value distribution followed by the largest value within the sample.
Nothing I present here is new.  However, from experience, I have found that a simple, short and compact guide on this matter written for the physics community is missing.   
\end{abstract}
\maketitle
\section{Introduction}
\label{intro}

Extreme value statistics offers a powerful tool box for the theoretical physicist.  But it is the kind of tool box that is not missed before one has been introduced
to it --- perhaps a little like the smart phone. It concerns the statistics of extreme events and it aims to answer questions like ``if the strongest signal I have observed
over the last hour had the value $x$, what would the strongest signal expected to be if measured over hundred hours?"  Furthermore, if I divide up this hundred-hour interval into a hundred
one-hour intervals, what would be the statistical distribution of strongest signal in each one-hour interval?  

It is the latter question which is the focus of this mini-review.  

There is no lack of literature on extreme value statistics, see e.g. \cite{g58,d81,g87,ekm97,c01} or simply {\it google\/} the term.  We find it used in connection with spin glasses
and disordered systems \cite{bm97}, in connection with $1/f$ noise \cite{adgr01}, in connection with optics \cite{rs12}, in connection with the fiber bundle model \cite{hhp15}, etc.
There are plenty of examples from diverse fields of physics.

So, there is no lack of material for the novice that has seen a need for this tool. The problem is that it is not so easy to penetrate the literature, which is often 
cast in a rather mathematical language which takes work to penetrate.  The aim of this mini-review is to present the theory behind and
the main results concerning the extreme value distributions in a simple and compact way.  We will present nothing new.   For a longer, wider and more detailed review of extreme value 
statistics, Fortin and Clusel \cite{fc15} or Majumdar et al.\ present exactly that \cite{mps20}.  
          
We have a statistical distribution $p(x)$ and its associated cumulative probability
\begin{equation}
\label{eq1}
P(x)=\int_{-\infty}^x p(x')dx'\;,
\end{equation}
which is the probability to find a number smaller than or equal to $x$.
We draw $N$ numbers from this distribution and record the largest of the $N$ numbers.  We repeat this procedure $M$ times and thereby obtain $M$ largest numbers,
one for each sequence.  What is the distribution of these $M$ largest
numbers in the limit when $M\to\infty$, which then defines the {\it extreme value distribution?\/}

It turns out that depending on $p(x)$, the extreme value distribution will have one of three functional forms:
\begin{itemize}
\item The {\it Weibull\/} cumulative probability
\begin{equation}
\Phi(u)=\left\{
\begin{array}{ll}
e^{-(-u)^\alpha} & {\rm for}\  u <   0\;,\\
1                & {\rm for}\  u \ge 0\;,\\
\end{array}
\right.
\label{eq3}
\end{equation} 
where we assume $\alpha > 0$. Note that $\Phi(-\infty)=0$.  
The corresponding Weibull extreme value distribution is 
\begin{equation}
\label{eq3-1}
\phi(u)=\left\{
\begin{array}{ll}
\alpha (-u)^{\alpha-1}e^{-(-u)^\alpha} & {\rm for}\  u <   0\;,\\
0                                      & {\rm for}\  u \ge 0\;.\\
\end{array}
\right.
\end{equation}
\item The {\it Fr{\'e}chet\/} cumulative probability
\begin{equation}
\Phi(u)=\left\{
\begin{array}{ll}
0                & {\rm for}\  u \le 0\;,\\
e^{-u^{-\alpha}} & {\rm for}\  u >   0\;.\\
\end{array}
\right.
\label{eq4}
\end{equation} 
Also here we assume $\alpha>0$. Note that $\Phi(\infty)=1$. The Fr{\'e}chet extreme value distribution is 
\begin{equation}
\label{eq4-1}
\phi(u)=\left\{
\begin{array}{ll}
0                                      & {\rm for}\  u \le 0\;,\\
\alpha u^{-\alpha-1}e^{-u^{-\alpha}}   & {\rm for}\  u >   0\;.\\
\end{array}
\right.
\end{equation}
\item The {\it Gumbel\/} cumulative probability
\begin{equation}
\label{eq2}
\Phi(u)=e^{-e^{-u}}\;,
\end{equation}
where $-\infty<u<\infty$, so that $\Phi(-\infty)=0$ and $\Phi(\infty)=1$. The corresponding Gumbel
extreme value distribution is given by
\begin{equation}
\label{eq2-1}
\phi(u)=e^{-u-e^{-u}}\;.
\end{equation}
\end{itemize}

The questions are 1.\ which classes of distributions $p(x)$ lead to which of the three extreme value distributions and 2.\ what is the connection between $x$ and $u$ in each case?  It turns out
that 
\begin{itemize}
\item distributions where $p(x)=0$ for $x>x_0$ and $p(x)\sim (x_0-x)^{\alpha-1}$ as $x\to x_0^-$, see equation (\ref{eq6-11}), lead to the {\it Weibull extreme value distribution,\/} 
\item distributions where $p(x)\sim x^{-\alpha-1}$ as $x\to \infty$, see equation (\ref{eq17-1}) lead 
to the {\it Fr{\'e}chet extreme value distribution,\/} 
\item and distributions where $p(x)$ falls of faster than any power law as $x\to \infty$, see equation (\ref{eq30-8}) lead to the {\it Gumbel extreme value distribution.\/}   
\end{itemize} 

Furthermore, we will find that
\begin{itemize}
\item for the {\it Weibull extreme value distribution,\/} $u$ is given in terms of $x$ in  equation (\ref{eq9-1}),  
\item for the {\it Fr{\'e}chet extreme value distribution,\/} $u$ given in terms of $x$ in equation (\ref{eq19}),
\item for the {\it Gumbel extreme value distribution,\/} $u$ is given in terms of $x$ in equations (\ref{eq30-6}) and (\ref{eq30}).  
\end{itemize} 

The discussion that will now follow, will be built on the following relation. When drawing $N$ numbers from the probability distribution $p(x)$, the cumulative probability
for the largest value is the probability that all $N$ values are either smaller than or equal to it. This probability is $P(x)^N$.  Our task is to figure out the asymptotic 
shape of $P(x)^N\to \Phi(u)$ as $N\to\infty$, and what is $u=u(x)$ as we approach this limit.  

\section{Weibull Class}
\label{weibull}

We consider here probability distributions $p(x)$ that have two characteristics:
\begin{itemize}
\item The associated cumulative probability $P(x)$ obeys
\begin{equation}
\label{eq6-11}
P(x_0)=1\quad{\rm and}\quad P(x_0-\Delta x)< 1\ {\rm for\ all}\ \Delta x>0\;.
\end{equation}
\item and it obeys   
\begin{equation}
\label{eq6-10}
\lim_{\Delta x\to 0^+}\frac{1-P(x_0-k\Delta x)}{1-P(x_0-\Delta x)}=k^\alpha\;,
\end{equation} 
where $k>0$ and $\alpha > 0$. 
\end{itemize}
This is equivalent to $p(x)$ having the form  
\begin{equation}
p(x)=\left\{
\begin{array}{ll}
b\alpha(x_0-x)^{\alpha-1} & {\rm for}\ x\to x_0^-\;,\\
0 & {\rm for}\ x > x_0\;,\\
\end{array}
\right.
\label{eq6-1}
\end{equation} 
where $b$ is positive. We note that $0 < \alpha < 1$ leads to
a diverging probability density as $x\to x_0^-$. We furthermore note
that $\alpha=1$ implies that $p(x)$ approach a constant when $x\to x_0^-$ ---
which for example is the case when the distribution is uniform. The corresponding 
cumulative probability is given by  
\begin{equation}
P(x)=\left\{
\begin{array}{ll}
1              & {\rm for}\  x \ge x_0\;,\\
1-b(x_0-x)^\alpha & {\rm for}\  x \to x_0^-\;.\\
\end{array}
\right.
\label{eq7-11}
\end{equation} 

The extreme value cumulative probability for $N$ samplings is given by 
\begin{equation}
\label{eq8-1}
P^N(x)=[1-b(x_0-x)^\alpha]^N\;,
\end{equation}
for $x\to x_0^-$. We introduce the variable change
\begin{equation}
\label{eq9-1}
x-x_0=\frac{u}{(bN)^{1/\alpha}}\;.
\end{equation}
Equation (\ref{eq8-1}) then becomes
\begin{equation}
\label{eq8-2}
P^N(x)=\left[1-\frac{(-u)^\alpha}{N}\right]^N\;.
\end{equation}
In the limit of $N\to\infty$, this becomes
\begin{equation}
\label{eq14-1}
\Phi(u)=\lim_{N\to\infty}P^N(x)=e^{-(-u)^\alpha}\;,
\end{equation}
for negative $u$. Hence, we have that
\begin{equation}
\Phi(u)=\left\{
\begin{array}{ll}
e^{-(-u)^\alpha} & {\rm for}\  u <   0\;,\\
1                & {\rm for}\  u \ge 0\;,\\
\end{array}
\right.
\label{eq7-1}
\end{equation} 
which is the {\it Weibull cumulative probability,\/} valid for {\it all\/} values of $u$ even
though we only know the behavior of $p(x)$ close to $x_0$.  The Weibull probability density is given by
\begin{equation}
\label{eq15}
\phi(u)=\frac{d\Phi(u)}{du}=\left\{
\begin{array}{ll}
\alpha (-u)^{\alpha-1}e^{-(-u)^\alpha} & {\rm for}\  u <   0\;,\\
0                                      & {\rm for}\  u \ge 0\;.\\
\end{array}
\right.
\end{equation}

We note that the Weibull distribution resembles a stretched exponential.  This is correct for $\alpha <1$.  However,
$\alpha \ge 1$ is much more common in the wild.  

We express the Weibull cumulative probability in terms of the original variable $x$ using equation (\ref{eq9-1}),
\begin{equation}
\label{eq15.1}
\Phi(u)=\Phi\left((bN)^{1/\alpha}(x-x_0)\right)=e^{-Nb(x_0-x)^\alpha}=\tilde{\Phi}(x)\;.
\end{equation}
Hence, in terms of the original variable $x$, the Weibull extreme value distribution becomes
\begin{equation}
\label{eq15.2}
\tilde{\phi}(x)=\frac{d\tilde{\Phi}(x)}{dx}=Nb\alpha(-x)^{\alpha-1}e^{-Nb(x_0-x)^\alpha}\;.
\end{equation}

\subsection{Weibull: An Example}
\label{weibullex}

We now work out a concrete example.  Let us assume that 
$p(x)$ is given by
\begin{equation}
p(x)=\left\{
\begin{array}{lr}
0 & {\rm for}\ x < 0\;,\\
\alpha (1-x)^{\alpha-1} & {\rm for}\ 0\le x \le 1\;,\\
0 & {\rm for}\ x > 1\;,
\end{array}
\right.
\label{eq6}
\end{equation}  
i.e., $b=1$ and $x_0=1$ in equation (\ref{eq6-1}). The cumulative probability is then
\begin{equation}
P(x)=\left\{
\begin{array}{lr}
0 & {\rm for}\ x < 0\;,\\
1-(1-x)^\alpha & {\rm for}\ 0\le x \le 1\;,\\
1 & {\rm for}\ x > 1\;.
\end{array}
\right.
\label{eq7}
\end{equation}

\begin{figure}
\includegraphics[width=\linewidth]{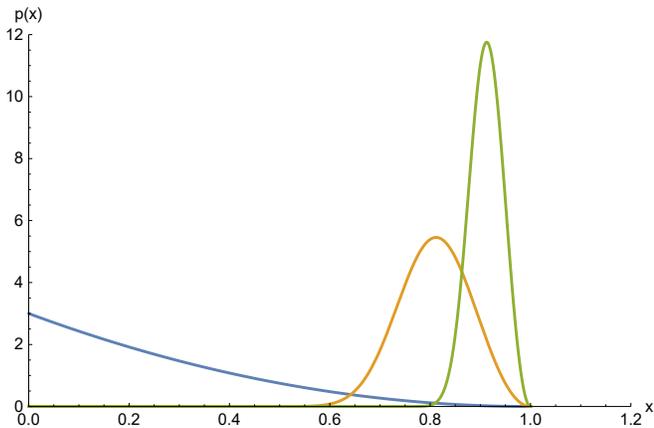}
  \caption{The curve that has its maximum at $x=0$ is the probability distribution (\ref{eq6}) with $\alpha=3$. The curve that has its maximum in the middle is $\tilde\phi(x)$, 
  equation (\ref{eq15.3}) with $N=100$ and the curve that has its maximum to the right is $\tilde\phi(x)$ with $N=1000$.} 
  \label{fig1}
\end{figure}

From equation (\ref{eq15.2}) and we have that 
\begin{equation}
\label{eq15.3}
\tilde{\phi}(x)=N\alpha(1-x)^{\alpha-1}e^{-N(1-x)^\alpha}\;.
\end{equation}
We show the distribution (\ref{eq6}) with $\alpha=3$ together with the corresponding extreme value distributions for $N=100$ and $N=1000$,
equation (\ref{eq15.2}) in figure \ref{fig1}.      

\begin{figure}
\includegraphics[width=\linewidth]{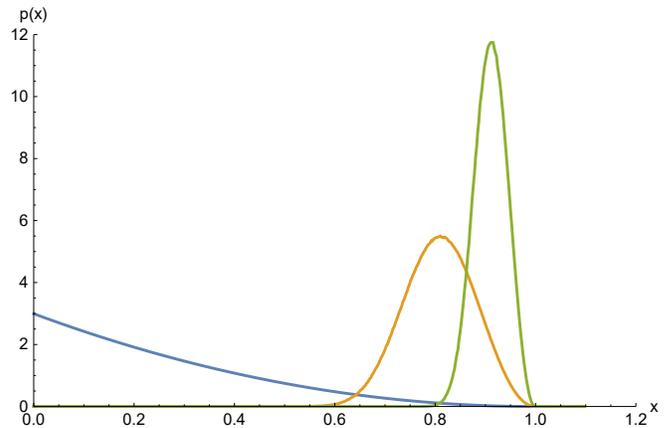}
  \caption{The histograms shown here are based on data according to the probability distribution (\ref{eq6}) with $\alpha=3$.  
           The histogram having its maximum to the left shows all the generated data.  The histogram having its maximum in the middle
           shows the largest number among each sequence of numbers of length 100, and the histogram having its maximum to the right
           shows the largest number among each sequence of numbers of length 1000. We generated $10^7$ sequences for both cases.}
  \label{fig2}
\end{figure}

Using a random number generator producing numbers $r$ uniformly distributed on the unit interval, we may stochastically generate numbers that 
are distributed according to the probability density $p(x)$ given in (\ref{eq6}).  We do this by inverting the expression $P(x)=r$, where the cumulative probability is
given by (\ref{eq7}).  Hence, we have
\begin{equation}
\label{eq15.4}
x=1-r^{1/\alpha}\;,
\end{equation}
where we have also used that $r$ may be substituted for $1-r$ in (\ref{eq7}).  We generate a sequence of sequences of numbers using this algorithm, each sequence having length $N$.
We then identify the largest value within each sequence.  We chose $N=100$ and $N=1000$, in each case generating $10^7$ such sequences.  The histograms based on the random numbers themselves, and of the extreme values for each sequence of length either 100 or 1000 we
show in figure \ref{fig2}.  This figure should be compared to figure \ref{fig1}.       

The Weibull distribution, equation (\ref{eq15}) is much used in connection with material strength \cite{r08}.  This is no coincidence. Consider a chain. Each link in the chain
can sustain a load up to a certain value, above which it fails.  This maximum value is distributed according to some probability distribution.  When the chain is loaded, it will be 
the link with the {\it smallest\/} failure threshold that will break first causing the chain as a whole to fail.  Hence, the strength distribution of an ensemble of chains
is an extreme value distribution, but with respect to the smallest rather than the largest value. The link strength must a positive number.  Hence, the link strength distribution is
cut off at zero or some positive value.  The distribution close to this cutoff value must behave as a power law in the distance to the cutoff, e.g.\ due to a Taylor expansion 
around the cutoff.  The corresponding extreme value distribution, which is the chain strength distribution, must then be a Weibull distribution.      

\section{Fr{\'e}chet Class}
\label{frechet}

We now assume that the probability distribution $p(x)$ whose associated cumulative probability behaves
as
\begin{equation}
\label{eq17-1}
\lim_{x\to\infty}\frac{1-P(x)}{1-P(kx)}=k^\alpha\;,
\end{equation}
where $k>0$ and $\alpha \ge 0$. This means that $p(x)$ behaves as
\begin{equation}
p(x)= b\alpha x^{-\alpha-1}\ {\rm for}\ x\to\infty\;,
\label{eq16}
\end{equation} 
and the corresponding cumulative probability behaves as
\begin{equation}
P(x)=1-b x^{-\alpha}\ {\rm for}\ x \to\infty\;.
\label{eq17}
\end{equation} 

The extreme value cumulative probability for $N$ samplings is given by 
\begin{equation}
\label{eq18}
P^N(x)=[1-bx^{-\alpha}]^N\;,
\end{equation}
for $x \to \infty$. We introduce the variable change
\begin{equation}
\label{eq19}
x=(bN)^{1/\alpha} u\;,
\end{equation}
We now plug this change of variables into equation (\ref{eq18}) to find
\begin{equation}
\label{eq23}
P^N(x)=\left[1-b\left((bN)^{1/\alpha}\ u\right)^{-\alpha}\right]^N=\left[1-\frac{u^{-\alpha}}{N}\right]^N\;.
\end{equation}
In the limit of $N\to\infty$, this becomes
\begin{equation}
\label{eq24}
\Phi(u)=\lim_{N\to\infty}P^N(x)=e^{-u^{-\alpha}}\;,
\end{equation}
where $u\ge 0$ is given by equation (\ref{eq19}). We see that $\Phi(u)\to 0$ as $u\to 0^+$. Furthermore, for $u<0$,
the function is no longer real.  Hence, we define $\Phi(u)=0$ for $u <0$.  The ensuing extreme value cumulative probability
is then given by  
\begin{equation}
\Phi(u)=\left\{
\begin{array}{ll}
0                & {\rm for}\  u \le   0\;,\\
e^{-u^{-\alpha}} & {\rm for}\  u > 0\;,\\
\end{array}
\right.
\label{eq24-1}
\end{equation} 
which is the {\it Fr{\'e}chet cumulative probability.\/} The Fr{\'e}chet probability density is given by
\begin{equation}
\label{eq25}
\phi(u)=\frac{d\Phi(u)}{du}=\left\{
\begin{array}{ll}
0                                      & {\rm for}\  u \le    0\;.\\
\alpha u^{-\alpha-1}e^{-u^{-\alpha}}   & {\rm for}\  u >  0\;.\\
\end{array}
\right.
\end{equation}

We express the Fr{\'e}chet cumulative probability in terms of the original variable $x$ using equation (\ref{eq19}),
\begin{equation}
\label{eq25.1}
\Phi(u)=\Phi\left(\frac{x}{(bN)^{1/\alpha}}\right)=e^{-Nx^{-\alpha}}=\tilde{\Phi}(x)\;.
\end{equation}
Hence, in terms of the original variable $x$, the Fr{\'e}chet extreme value distribution becomes
\begin{equation}
\label{eq25.2}
\tilde{\phi}(x)=\frac{d\tilde{\Phi}(x)}{dx}=N\alpha x^{-\alpha-1}e^{-Nx^{-\alpha}}\;.
\end{equation}

\subsection{Fr{\'e}chet: An Example}
\label{frechetex}

\begin{figure}
\includegraphics[width=\linewidth]{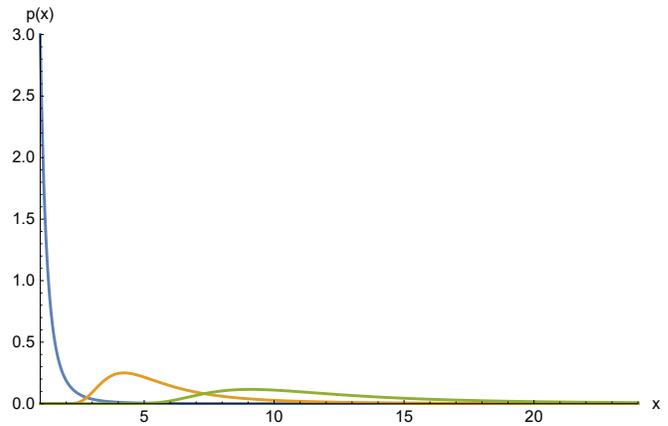}
 \caption{The curve that has its maximum at $x=1$ is the probability distribution (\ref{eq25.3}) with $\alpha=3$. The curve that has its maximum in the middle is $\tilde\phi(x)$, 
  equation (\ref{eq25.5}) with $N=100$ and the curve that has its maximum to the right is $\tilde\phi(x)$ with $N=1000$.} 
  \label{fig3}
\end{figure}

We consider the distribution 
\begin{equation}
p(x)=\left\{
\begin{array}{lr}
0 & {\rm for}\ x \le 1\;,\\
\alpha x^{-\alpha-1} & {\rm for}\ x > 1 \;,\\
\end{array}
\right.
\label{eq25.3}
\end{equation}
The corresponding cumulative probability is given by   
\begin{equation}
P(x)=\left\{
\begin{array}{lr}
0 & {\rm for}\ x \le 1\;,\\
1-x^{-\alpha} & {\rm for}\ x > 1\;.\\
\end{array}
\right.
\label{eq25.4}
\end{equation}
Using equation (\ref{eq25.2}), we find the corresponding Fr{\'e}chet extreme value distribution to be 
\begin{equation}
\label{eq25.5}
\tilde{\phi}(x)=N\alpha x^{-\alpha-1}e^{-Nx^{-\alpha}}\;,
\end{equation}
valid for all $x>1$.  We show $p(x)$ and the corresponding $\tilde\phi(x)$ for $\alpha=3$ and $N=100$ and
$N=1000$ in figure \ref{fig3}. 

\begin{figure}
\includegraphics[width=\linewidth]{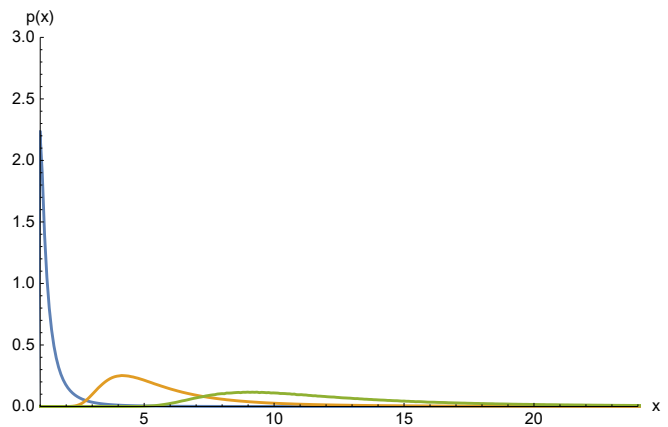}
  \caption{The histograms shown here are based on data according to the probability distribution (\ref{eq25.3}) with $\alpha=3$.  
           The histogram having its maximum to the left shows all the generated data.  The histogram having its maximum in the middle
           shows the largest number among each sequence of numbers of length 100, and the histogram having its maximum to the right
           shows the largest number among each sequence of numbers of length 1000.  For each sequence length, $10^7$ such sequences were generated.}
  \label{fig4}
\end{figure}

In order to compare with numerical results, we generate numbers distributed according to (\ref{eq25.3}) by solving the equation $P(x)=r$ where $r$ is drawn from
a uniform distribution on the unit interval.  From equation (\ref{eq25.4}), we get
\begin{equation}
\label{eq25.6}
x=r^{-1/\alpha}\;.
\end{equation} 
We generate a sequence of numbers using this algorithm, grouping them together in sequences of
$N=100$ or $N=1000$.  We generate $10^7$ such sequences.  The histograms based on the random numbers themselves generated with equation (\ref{eq25.6}), and of the extreme values for each sequence 
of length either 100 or 1000 we show in figure \ref{fig4}.  This figure should be compared to figure \ref{fig3}.   

\section{Gumbel Class}
\label{gumbel}

We now assume we have a probability distribution that takes the form
\begin{equation}
\label{eq26}
p(x)=f'(x) e^{-f(x)}\quad {\rm for}\ x > x_0 \;,
\end{equation}
where $f'(x)=df(x)/dx$.  We have that $x_0$ is any number, positive or negative, and 
$f(x)$ is an increasing function with $x$. The cumulative probability is then
\begin{equation}
P(x)=1-e^{-f(x)}\quad{\rm for}\ x > x_0 \;.
\label{eq27}
\end{equation}
We do not care about the form of $p(x)$ or $P(x)$ for $x\le x_0$. 

The extreme value cumulative probability for $N$ samplings is given by 
\begin{equation}
\label{eq28}
P^N(x)=[1-e^{-f(x)}]^N\;,
\end{equation}
for $x > x_0$. We introduce the variable change
\begin{equation}
\label{eq29}
\tilde{u}=f(x)-f(x_N)\;,
\end{equation}
where $x_N$ is given by  
\begin{equation}
\label{eq30}
P(x_N)=1-\frac{1}{N}\;.
\end{equation}
From equation (\ref{eq27}) we then have that
\begin{equation}
\label{eq31}
f(x_N)=\ln N\;.
\end{equation}

Let us now define
\begin{equation}
\label{eq30-1}
\Delta x=x-x_N\;.
\end{equation}
We then expand $f(x)$ around $x_N$,
\begin{equation}
\label{eq30-2}
f(x)=f(x_N+\Delta x)=\sum_{n=0}^\infty \frac{f^{(n)}(x_N)}{n!}\Delta x^n\;,
\end{equation}
where $f^{(n)}(x)=d^nf(x)/dx^n$. If we now set 
\begin{equation}
\label{eq30-3}
\Delta x = \frac{1}{f'(x_N)}\;,
\end{equation}
so that the first order term in the expansion becomes constant as $N$ increases, we will have
that
\begin{equation}
f'(x_N)\Delta x +\sum_{n=2}^\infty \frac{f^{(n)}(x_N)}{n!}\Delta x^n
=1+\sum_{n=2}^\infty \frac{f^{(n)}(x_N)}{n!f'(x_N)^n}\;.
\end{equation}
Hence, if we have that 
\begin{equation}
\label{eq30-4}
\lim_{N\to\infty}\frac{f^{(n)}(x_N)}{f'(x_N)^n}=0\;,
\end{equation}
then in this limit, we will find
\begin{equation}
\label{eq30-5}
f(x)=f(x_N)+f'(x_N)\Delta x=f(x_N)+u\;,
\end{equation}
where we define
\begin{equation}
\label{eq30-6}
u=f'(x_N)\Delta x=Np(x_N)(x-x_N)\;.
\end{equation}
Here we have used equations (\ref{eq27}) and (\ref{eq31}). 

If we combine equation (\ref{eq30-4}) for $n=2$ with equations (\ref{eq26}) and (\ref{eq27}), we find
\begin{equation}
\label{eq30-7}
\lim_{N\to\infty}\frac{f''(x_N)}{f'(x_N)^2}=\lim_{N\to\infty}\frac{d}{dx}\left[\frac{1-P(x)}{p(x)}\right]_{x=x_N}=0\;,
\end{equation}
which is equivalent to
\begin{equation}
\label{eq30-8}
\lim_{x\to\infty}\frac{d}{dx}\left[\frac{1-P(x)}{p(x)}\right]=0\;.
\end{equation}
Equation (\ref{eq30-8}) is in fact a {\it sufficient condition\/} for (\ref{eq30-4}) to hold for all $n > 1$. We may show 
this through induction.  We have that
\begin{equation}
\label{eq30-9}
\frac{f^{(n+1)}(x)}{f'(x)^{n+1}}=\frac{1}{f'(x)}\ \frac{d}{dx}\left(\frac{f^{(n)}(x)}{f'(x)^n}\right)+\frac{f^{(n)}(x)}{f'(x)^{n+2}}\;.
\end{equation} 
If condition (\ref{eq30-7}) is fulfilled, that is when the expression above is zero in the limit $x\to\infty$ for $n=2$, we also have that 
\begin{equation}
\label{eq30-10}
\lim_{N\to\infty}\frac{f^{(3)}(x)}{f'(x)^{3}}=0\;,
\end{equation}
since both terms on the right hand side of equation (\ref{eq30-9}) are zero in this limit.  We now assume equation (\ref{eq30-4}) to
be true for some $n>3$.  We then have that 
\begin{equation}
\label{eq30-11}
\lim_{N\to\infty}\frac{f^{(n+1)}(x_N)}{f'(x_N)^{n+1}}=0\;,
\end{equation}
again due to both terms on the right hand side of equation (\ref{eq30-9}) are zero in this limit. This completes the proof.

We now combine equations (\ref{eq29}) with equation (\ref{eq28}) to find  
\begin{eqnarray}
\label{eq33}
P^N(x)&=&\left[1-e^{-u-f(x_N)}\right]^N\nonumber\\
      &=&\left[1-e^{-u-\ln N}\right]^N=\left[1-\frac{e^{-u}}{N}\right]^N\;.
\end{eqnarray}
In the limit of $N\to\infty$, this becomes
\begin{equation}
\label{eq34}
\Phi(u)=\lim_{N\to\infty}P^N(x)=e^{-e^{- u}}\;,
\end{equation}
which is the {\it Gumbel cumulative probability.\/} Here $-\infty < u < \infty$. The Gumbel probability density is given by
\begin{equation}
\label{eq35}
\phi(u)=\frac{d\Phi(u)}{du}=e^{-u-e^{-u}}\;.
\end{equation}

We express the Gumbel cumulative probability in terms of the original variable $x$ using equation (\ref{eq30-6}),
\begin{eqnarray}
\label{eq35.1}
\Phi(u)&=&\Phi\left(Np(x_N)(x-x_N)\right)\nonumber\\
       &=&e^{-e^{-Np(x_N)(x-x_N)}}=\tilde{\Phi}(x)\;.
\end{eqnarray}
Hence, in terms of the original variable $x$, the Gumbel extreme value distribution becomes
\begin{eqnarray}
\label{eq35.2}
\tilde{\phi}(x)&=&\frac{d\tilde{\Phi}(x)}{dx}\nonumber\\
               &=&Np(x_N)e^{-Np(x_N)(x-x_N)-e^{-Np(x_N)(x-x_N)}}\;.\nonumber\\
\end{eqnarray}

\subsection{An Example: the Gaussian}
\label{gauss}

Here is an example: the gaussian.  The gaussian probability density is given by
\begin{equation}
\label{eq36}
p(x)=\frac{e^{-x^2/2\sigma}}{\sqrt{2\pi \sigma}}\;,
\end{equation}
where $\sigma$ is the square of the standard deviation.  The cumulative probability is 
\begin{equation}
\label{eq37}
P(x)=\frac{1}{2}\left[1+{\rm erf}\left(\frac{x}{\sqrt{2\sigma}}\right)\right]\;,
\end{equation}
where ${\rm erf}(x)$ is the error function. In order to verify that the gaussian generates the Gumbel extreme
distribution, we use the sufficient condition (\ref{eq30-8}),
\begin{eqnarray}
\label{eq38}
&&\lim_{x\to\infty}\frac{d}{dx}\left[\frac{1-P(x)}{p(x)}\right]=\nonumber\\
&=&\lim_{x\to\infty}\sqrt{\frac{\pi}{2\sigma}}e^{x^2/2\sigma}x\left[1-{\rm erf}\left(\frac{x}{\sqrt{2\sigma}}\right)\right]=0\;.
\end{eqnarray}

The gaussian cumulative probability in equation (\ref{eq37}) has the asymptotic form
\begin{equation}
\label{eq39}
P(x)=1-\sqrt{\frac{\sigma}{2\pi}}\frac{e^{-x^2/2\sigma}}{x}\;,
\end{equation}
for large $x$. We determine $x_N$ solving equation (\ref{eq30}) using this asymptotic form.  We find
\begin{equation}
\label{eq40}
x_N=\sqrt{\sigma W\left(\frac{N^2}{2\pi}\right)}\;,
\end{equation}
where $W(z)$ is the Lambert W function, also known as the product logarithm, which is the solution to the equation $W(z)\exp[W(z)]=z$. For large 
arguments, it approaches the natural logarithm. This gives us
\begin{equation}
\label{eq41}
Np(x_N)=\sqrt{\frac{1}{\sigma} W\left(\frac{N^2}{2\pi}\right)}\;,
\end{equation}
when inserting the expression for $x=x_N$, equation (\ref{eq40}) into equation (\ref{eq36}).  Thus we may now express the variable $u$ in the Gumbel 
cumulative probability (\ref{eq33}) in terms of the variables $x$, $\sigma$ and $N$ using equation (\ref{eq30-6}),
\begin{equation}
\label{eq42}
u=x\ \sqrt{\frac{1}{\sigma} W\left(\frac{N^2}{2\pi}\right)} - W\left(\frac{N^2}{2\pi}\right)\;.
\end{equation}

\begin{figure}
\includegraphics[width=\linewidth]{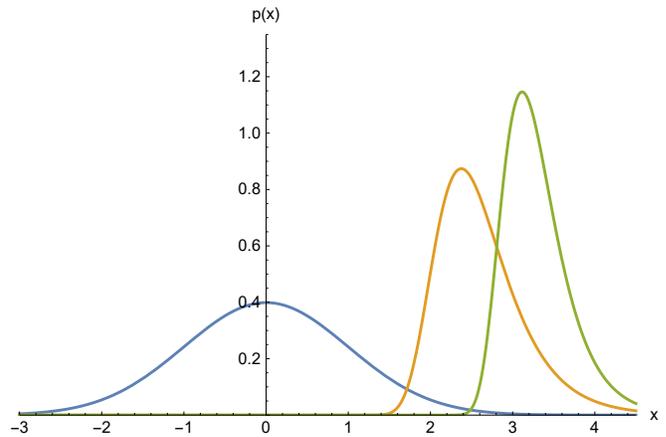}
  \caption{The gaussian and the corresponding Gumbel distributions for $\sigma=1$ and $N=100$ and $N=1000$.}
  \label{fig5}
\end{figure}

We show in figure \ref{fig5} the gaussian and the corresponding Gumbel distributions for $\sigma=1$ and $N=100$ and $N=1000$. We find that $x_{100}=2.375$ and
$x_{1000}=3.115$. These are the confidence intervals for 99\% and 99.9\%.
 
\begin{figure}
\includegraphics[width=\linewidth]{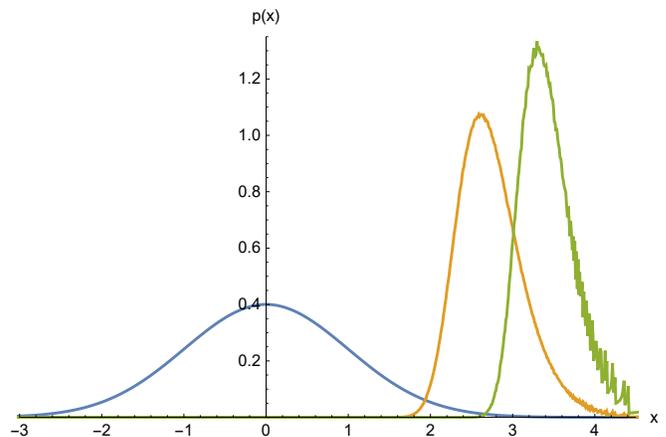}
  \caption{The histograms shown here are based on data generated using the Box-M{\"u}ller algorithm which produces numbers distributed according to a 
           gaussian.  Here $\sigma=1$.  The histogram with the maximum to the left shows all the generated data.  The histogram with its maximum in the middle
           shows the largest number among each sequence of numbers of length 100, and the histogram with the rightmost maximum
           shows the largest number among each sequence of numbers of length 1000.  For each sequence length, $10^7$ such sequences were generated.}
  \label{fig6}
\end{figure}

We show in figure \ref{fig6} a histogram based on numbers distributed according to a gaussian distribution using the Box-M{\"u}ller algorithm \cite{ptvf07}.  
These numbers were grouped together in sets of either $N=100$ or $N=1000$ elements. I generated $10^7$ such sets.  The figure displays the two extreme distributions
for the two set sizes.  This figure should be compared to figure \ref{fig5}.  In contrast to the two other extreme value distributions, we see that there are visible 
discrepancies between the calculated Gumbel distributions in figure \ref{fig5} and the extreme value histograms in figure \ref{fig6}.  We see furthermore that the histogram 
for $N=1000$ is closer to the calculated Gumbel distribution than the histogram for $N=100$.  This is due to the very slow convergence induced by the Lambert W functions.    
Slow convergence is typical for the Gumbel extreme value distributions. This slow convergence has been analyzed and recently and through clever use of scaling methods remedied
\cite{zbk20}.  
   
\section{Concluding Remarks}
\label{conclusion}

We have only discussed the distributions associated with the {\it largest\/} values of $x$ except for the Weibull extreme value distribution, section \ref{weibull}.  It is, however, easy to work out: 
{\it just transform $x\to -x$.\/}  Otherwise, the story presented here is rather complete.  

There is one remark that needs to be made, though. In the derivation of the Gumbel extreme value distribution, section \ref{gumbel}, we defined a variable $x_N$ in equation (\ref{eq30}).  First of all, $x_N$ 
defined in equation (\ref{eq30}) may be calculated for {\it any\/} cumulative probability $P(x)$ and it has an interpretation making it very useful.

The probability density for the largest among $N$ numbers drawn using the probability distribution $p(x)$ is given by
\begin{equation}
\label{eq40}
p_N(x)=\frac{dP(x)^N}{dx}=NP(x)^{N-1}p(x)\;.
\end{equation}
We calculate the average of the cumulative probability $P(x)$ for the extreme value based on $N$ samples,
\begin{eqnarray}
\label{eq41}
\langle P(x)\rangle&=&\int_{-\infty}^{\infty} P(x) p_N(x) dx\nonumber\\
&=&\int_0^1 P^N dP=\frac{N}{N+1}=1-\frac{1}{N+1}\;.
\end{eqnarray}
For large $N$, we may write this as
\begin{equation}
\label{eq42}
\langle P(x)\rangle=P(x_N)=1-\frac{1}{N}\;,
\end{equation}
using here equation (\ref{eq30}).    
Hence, we may interpret $x_N$ as the $x$ value corresponding to the average confidence interval of the largest observed value in sequences of $N$ numbers.  It is essentially the typical size of the
extreme value for a sample of size $N$.   

\begin{acknowledgements}
I thank Eivind Bering, Astrid de Wijn, H.\ George E.\ Hentschel, Srutarshi Pradhan and 
Itamar Procaccia for numerous interesting discussions on this topic. This work was partly supported by the
Research Council of Norway through its Centers of Excellence funding
scheme, project number 262644. 
\end{acknowledgements}



\begin{thebibliography}{10}


\bibitem{g58} E. J. Gumbel, {\sl Statistics of Extremes\/} (Columbia University Press, New York, 1958).

\bibitem{d81} H. A. David, {\sl Order Statistics,\/} second ed.\ (Wiley, New York, 1981).

\bibitem{g87} J. Galambos, {\sl The Asymptotic Theory of Extreme Order Statistics\/} (Krieger, Malabar, FL, 1987).

\bibitem{ekm97} P.\ Embrechts, C.\ Kl{\"u}ppelberg and T.\ Mikosh, {\it Modeling Extreme Events for Insurance and Finance\/} (Springer, Berlin, 1997).

\bibitem{c01} S.\ Coles, {\it An Introduction to Statistical Modeling of Extreme Events\/} (Springer, Berlin, 2001). 

\bibitem{bm97} J. -P. Bouchaud and M. Mezard,Universality Classes for Extreme-Value Statistics, J. Phys. A, {\bf 30}, 7997 (1997). doi.org/10.1088/0305-4470/30/23/004.

\bibitem{adgr01} T. Antal, M. Droz, G. Gy{\"o}rgyi, and Z. R{\'a}cz, 1/f Noise and Extreme Value Statistics, Phys. Rev. Lett. {\bf 87}, 240601 (2001).
https://doi.org/10.1103/PhysRevLett.87.240601.

\bibitem{rs12} S. Randoux and P. Suret, Experimental Evidence of Extreme Value Statistics in Raman Fiber Lasers, Optics Lett. {\bf 37}, 500 (2012). doi.org/10.1364/OL.37.000500.   

\bibitem{hhp15} A. Hansen, P. C. Hemmer and S. Pradhan, {\sl The Fiber Bundle Model\/} (Wiley-VCH, Berlin, 2015).

\bibitem{fc15} J.-Y.\ Fortin and M.\ Clusel, {\it Applications of Extreme Value in Physics,\/} J.\ Phys.\ A: Math.\ Theor.\ {\bf 48}, 183001 (2015).
doi:10.1088/1751-8113/48/18/183001 

\bibitem{mps20} S. N. Majumdar, A. Pal and G. Schehr, Extreme Value Statistics of Correlated Random Variables: A Pedagogical Review, Phys. Rep., {\bf 840}, 1 (2020). 
doi.org/10.1016/j.physrep.2019.10.005 



\bibitem{ptvf07} W. H. Press, S. A. Teukolsky, W. T. Vetterling and B. P. Flannery, {\sl Numerical Recipes,\/} third ed.\ (Cambridge University Press, Cambridge, 2007).

\bibitem{r08} H. Rinne, {\sl The Weibull Distribution\/} (CRC Press, Boca Raton, 2008).

\bibitem{zbk20} L.\ Zarfaty, E.\ Barkai and D.\ A.\ Kessler, {\it Accurately Approximating Extreme Value Statistics,\/} arXiv:2006.13677 (2020).  


\end{thebibliography}
\end{document}